\documentclass[twocolumn,showpacs,%
  nofootinbib,aps,superscriptaddress,%
  eqsecnum,prd,notitlepage,showkeys,10pt]{revtex4-1}
\usepackage{float}
\makeatletter
\let\newfloat\newfloat@ltx
\makeatother
\usepackage{amssymb}
\usepackage{amsmath}
\usepackage{graphicx}
\usepackage{dcolumn}
\usepackage{hyperref}
\usepackage{amsfonts}
\usepackage{algpseudocode}
\usepackage{textcomp}
\usepackage{xcolor}
\usepackage{braket}
\usepackage{algorithm}
\usepackage{amsmath}
\usepackage{algorithmicx}

\begin{document}

\title{Non-Local Phase Estimation with a Rydberg-Superconducting Qubit Hybrid}
\author{Juan C. Boschero}
\affiliation{Applied Cryptography and Quantum Algorithms, TNO}
\author{Niels M.P. Neumann}
\affiliation{Applied Cryptography and Quantum Algorithms, TNO}
\affiliation{Algorithms and Complexity, Centrum Wiskunde \& Informatica}
\author{Ward van der Schoot}
\affiliation{Applied Cryptography and Quantum Algorithms, TNO}
\author{Frank Phillipson}
\affiliation{Applied Cryptography and Quantum Algorithms, TNO}
\affiliation{School of Business and Economics, Maastricht University}

\begin{abstract}
Distributed quantum computing (DQC) is crucial for high-volume quantum processing in the NISQ era. Many different technologies are utilized to implement a quantum computer, each with a different advantages and disadvantages. Various research is performed on how to implement DQC within a certain technology, but research on DQC between different technologies is rather limited. In this work, we contribute to this latter research line, by implementing the Quantum Phase Estimation algorithm on a superconducting-resonator-atom hybrid system. This system combines a Rydberg atom qubit, as well as a superconducting flux qubit system to perform the algorithm. In addition, Hamiltonian dynamics are studied to analyze noise sources, after which quantum optimal control (GRAPE) is used to optimize gate construction. The results show tradeoffs between GRAPE step size, iterations and noise level. \\ \\

\end{abstract}

\maketitle

\section{Introduction}

Quantum computing has garnered immense interest due to its potential to revolutionize information processing by harnessing quantum phenomena such as superposition and entanglement. As the field advances, the major challenge to scale quantum computers to perform increasingly complex tasks remains. Currently, quantum computing finds itself in the noisy intermediate-scale (NISQ) era characterized by quantum processors containing up to 1000 noisy qubits \cite{Preskill_2018, Brooks2019BeyondQS}. One promising solution is distributed quantum computing, where multiple quantum processors are interconnected to function as a larger, unified system \cite{Boschero2024DistributedQC}.
However, quantum computers are heterogeneous: Various quantum computing architectures have been theorized or developed, each offering unique advantages \cite{Noel_2022,Struck2016,Lachance_Quirion_2019,Psaroudaki_2021}. For example, superconducting flux qubits enable fast operation and integration with conventional electronics \cite{orlando_flux}, while Rydberg atom systems provide excellent coherence properties and precise atomic-level control \cite{ising_rydberg}. By combining these disparate technologies in a distributed system, one can exploit their individual strengths and mitigate inherent limitations.

In this research, we consider the phase estimation algorithm, a sub-process in for instance Shor's algorithm \cite{shor_1994} and the Harrow-Hassidim-Lloyd (HHL) quantum algorithm \cite{Zaman_2023}, and numerically simulate a distributed version of it on a system consisting of a flux qubit-based quantum computer and a Rydberg atom quantum computer. The quantum phase estimation algorithm determines the phase $\varphi$ of an unknown eigenvalue $e^{2\pi i \varphi}$ for a unitary operator $U$ with eigenvector $\ket{u}$ \cite{nielsen_chuang_2021}.

This hybrid approach explores the integration of different quantum architectures, with the goal of enabling scalable and fault-tolerant quantum computing networks. Thus, \textit{non-local} quantum computing, as discussed in this paper, refers to quantum computers that are spatially separated yet collaboratively executing an operation or algorithm. The numerical simulation builds upon the coupled system described in \cite{Yu_atom_flux_2017} to establish a connection between the two quantum computers, extending the system by incorporating more qubits and robust quantum models. First, the quantum systems and their corresponding noise models are derived. To optimize the system's evolution while accounting for noise, the Gradient Ascent Pulse Engineering (GRAPE) method \cite{KHANEJA2005296} is applied within the distributed phase estimation algorithm. The application and demonstration of the GRAPE algorithm address potential software engineering solutions and future challenges in error correction and optimal control.
This paper is based on the research presented in \cite{juan_supp}.

This work brings together three different research lines, namely distributed quantum computing, heterogeneous hybrid quantum computing, and quantum optimal control theory. For the first, earlier work has focused on theoretical descriptions of distributed quantum algorithms, examples being~\cite{Yimsiriwattana_2004, niels_imperfect, GHZ_distributed, Neumann_2022}, and physical experiments of homogeneous distributed quantum systems~\cite{photon_distributed, LaRacuente_2025}. For the second, earlier work has focused on theoretical descriptions and experimental implementations, with most notable works from ~\cite{werschnik2007quantumoptimalcontroltheory, Krotov, Doria_2011} to optimize gates. This work is the first to showcase the potential of combining not just any two of these research lines, but even all three of them.

This paper is structured as follows. First, an introduction of distributed quantum computing is given through a distributed version of the Quantum Phase Estimation algorithm in Section 2. Then, the hybrid atom-flux system is modeled as an open quantum system in Section 3. This model is used in Section 4 to optimize the noise level of the quantum algorithm. Lastly, conclusions and directions for future research are written in Section 5. 

\section{The Non-Local Phase Estimation Algorithm}
This section details a distributed version of the quantum phase estimation algorithm, and hence also gives an introduction into distributed quantum computing algorithms.

\subsection{The Phase Estimation Algorithm}
The phase estimation algorithm uses two registers of $n$ and $m$ qubits, respectively, called the counting and state qubits, respectively. Let $U$ be a unitary operator acting on the $m$-qubit register. The eigenvalues of a unitary operator have unit modulus and are characterized by their phase. If $\ket{\psi}$ is an eigenvector of $U$, then,
\begin{equation}
U \ket{\psi} = e^{2\pi i \theta} \ket{\psi}
\end{equation}  
for some $\theta \in \mathbb{R}$. Due to the periodicity of the complex exponential, we assume $0 \leq \theta < 1$. The goal is to efficiently approximate $\theta$. The quantum phase estimation algorithm accomplishes this by assuming oracular access to $U$ and availability of $\ket{\psi}$ as a quantum state. The algorithm returns an approximation of $\theta$ with additive error $\varepsilon$ using $O(1/\varepsilon)$ controlled-$U$ operations, where $\varepsilon$ is determined via $n = O(\log(1/\varepsilon))$~\cite{nielsen_chuang_2021}.

First, the algorithm creates a uniform superposition in the first register and then applies a sequence of controlled phase rotations given by $U^k$ for $k = 0,...,2^n-1$. This gives the state
\begin{equation}\label{eq:phase_estimation_state}
\begin{split}
    &\frac{1}{2^{n/2}}(\ket{0}+e^{i2\pi 2^{n-1}\varphi}\ket{1})(\ket{0}+e^{2\pi 2^{n-2}\varphi}\ket{1})\cdots \\
    &(\ket{0}+e^{2 \pi i 2^0\varphi}\ket{1})\ket{\psi} = \frac{1}{2^{n/2}}\sum^{2^n - 1}_{k=0}e^{2\pi i\varphi k}\ket{k}\ket{\psi}.
\end{split}
\end{equation}
An inverse Fourier transform on the first register then extracts this phase $\phi$ and produces the state $\ket{\varphi_1...\varphi_n}$.
Measuring in the computational basis returns $\varphi$ up to additive error $\varepsilon$~\cite{nielsen_chuang_2021}.

\subsection{Distributed Quantum Algorithms}
In distributed quantum systems, the qubits are distributed over different quantum processors. Because of this, interactions between certain pairs of qubits cannot be directly applied. Interactions between these qubits are called non-local (as opposed to local qubit operations), and are the basis building block of distributed quantum algorithms. To implement these non-local operations, so-called channel qubits need to be sent back and forth over a quantum network~\cite{GHZ_distributed}.

Given a normal quantum algorithm, it can be made distributed by replacing the local operations between qubits on different quantum processors, by non-local alternatives~\cite{Eisert_2000}. Implementing a distributed 2-qubit gate requires a shared entangled qubit pair. This is done by having the one processor send a channel qubit to the other machine. The second machine then entangles this qubit a local qubit. Finally, one qubit of the pair is sent back to the remote machine resulting in the shared entanglement \textit{GHZ state} $\frac{1}{2}(\ket{00}+\ket{11})$~\cite{GHZ_distributed}. This process of establishing entanglement between two different registers is called the entanglement gate, or equivalently, $E_2$ gate.~\cite{GHZ_distributed,Yimsiriwattana_2004,radzihovsky_espinosa_gim_2021,Eisert_2000}.

\begin{figure}[htbp]
\centerline{\includegraphics[width=\linewidth]{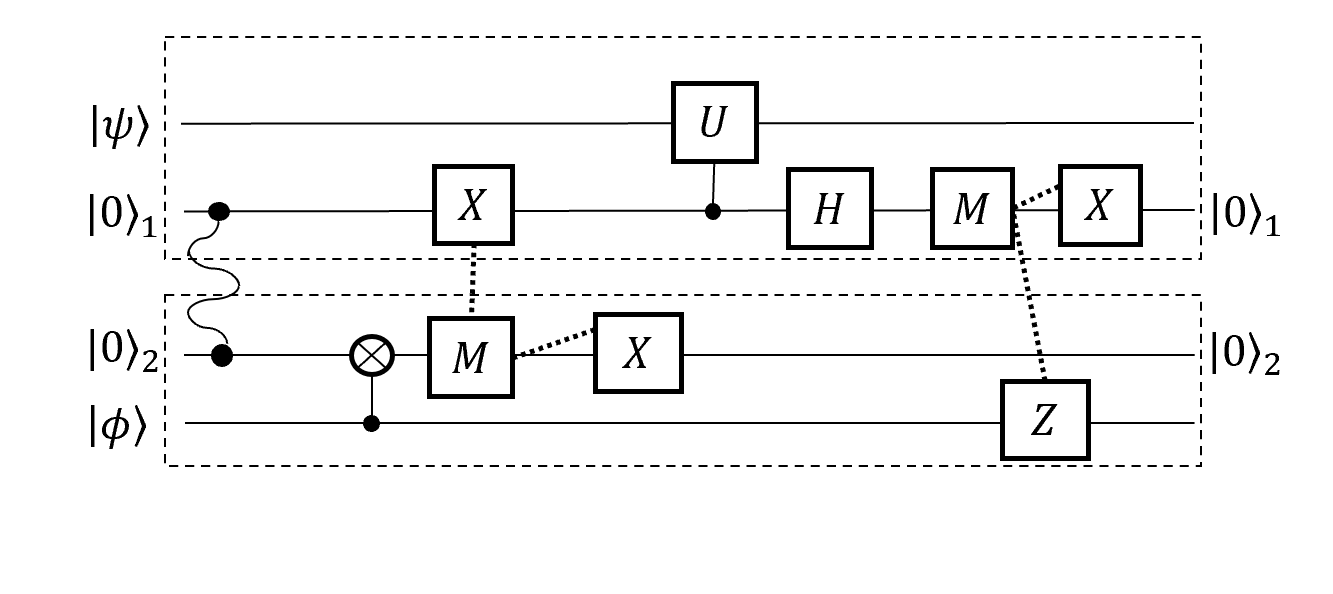}}
\caption{Implementation of a non-local controlled $U$-gate between $\ket{\phi}$ and $\ket{\psi}$ where the dotted box shows the different quantum computers and the dotted lines denote classical communication. The two dots with the wave in between denotes the $E_2$-gate \cite{niels_imperfect,Yimsiriwattana_2004,GHZ_distributed}.}
\label{fig: dis_u_gate}
\end{figure}

Now, suppose Alice and Bob, both holding their own quantum processor, want to implement a distributed $U$-gate between the state $\ket{\phi}$ at Alice's processor and $\ket{\psi}$ at Bob's processor. First, they apply an $E_2$ gate, resulting in two entangled qubits (ebits) at both devices. Alice then applies a CNOT gate between $\ket{\phi}$ and their ebit. Afterwards, she measures her ebit, and if the result is $\ket{1}$, she applies a NOT or $X$-gate to her ebit, and informs Bob about the measurement result. Bob then applies a NOT gate to his ebit if Alice's result was $\ket{1}$.
Bob can now use his ebit for local computations, applying a controlled $U$-gate between it and his qubit $\ket{\psi}$. 
To finalize the non-local CNOT-gate, Bob applies a Hadamard gate to his ebit, measures it, and, if the outcome is $\ket{1}$, applies a NOT gate and communicates the result to Alice. If Alice receives a $\ket{1}$, she applies a $Z$-gate to her qubit. These steps correspond to the non-local $U$-gate as shown in Fig. \ref{fig: dis_u_gate}~\cite{GHZ_distributed,Yimsiriwattana_2004,radzihovsky_espinosa_gim_2021}.

This study follows this protocol to implement the distributed phase estimation algorithm. One quantum computer holds the state qubits along with a single counting qubit, while the other device manages the other counting qubits to improve estimation accuracy. From the above, it can be concluded that the distributed phase algorithm rests on two key distributed processes: the distributed Quantum Fourier Transform~\cite{niels_imperfect} and the distributed controlled $U$-gate. 

\section{Modeling Hybrid System as Open Quantum System}
This section provides a high-level overview of the different parts in the hybrid system. 
More details and motivation for certain choices are found in literature, such as~\cite{juan_supp}.

\subsection{Rydberg Atom System}\label{sec:rydberg_atom_system}
\begin{figure}[t]
\centering
    \centerline{\includegraphics[width=0.5\textwidth]{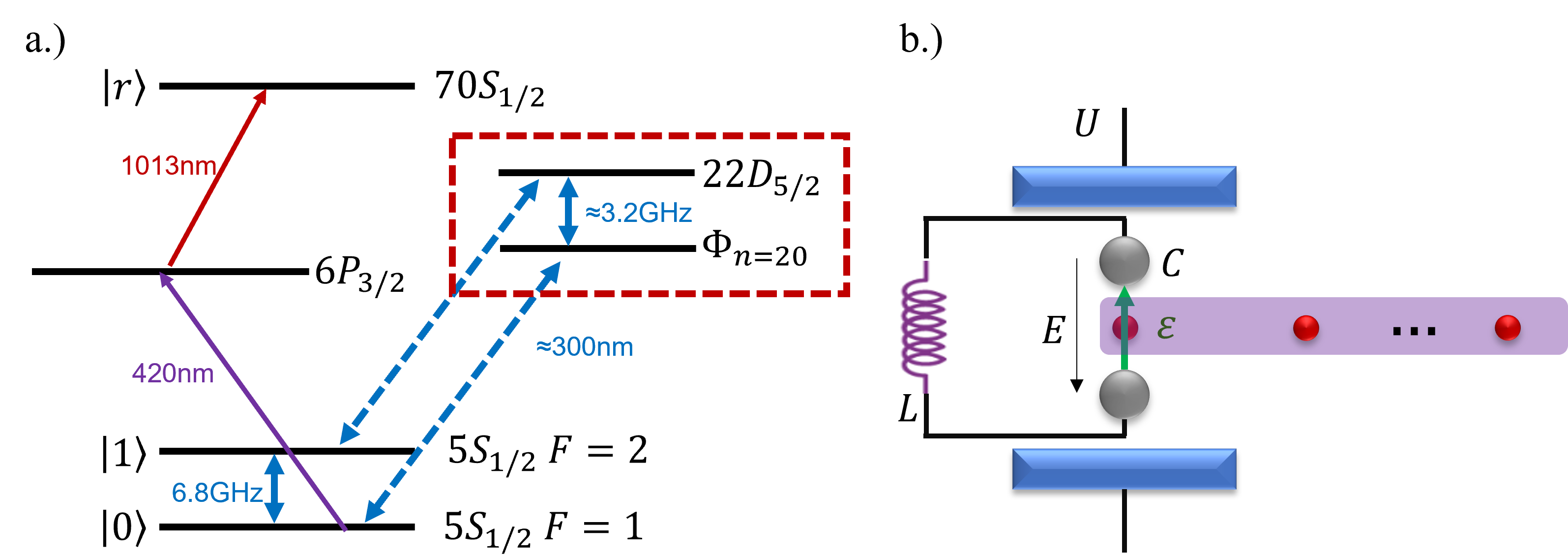}}
    \caption{(a) Schematic detailing the Rydberg energy states paired to the respective computational states of the proposed Rydberg computer atom array with the energy states of the communication qubit depicted in the red dotted square. The atoms in the atom array contain two computational levels in the hyperfine $\ket{5S_{1/2},F=1,2}$ states, the Rydberg auxiliary state in the $\ket{70S_{1/2}}$ state and an extra auxiliary $6P_{3/2}$ to facilitate the transfer to the $\ket{70S_{1/2}}$ state. The communication qubit in the red dotted box shows the two energy levels used in the work done in \cite{Yu_atom_flux_2017} whereby two energy states, $\ket{22D_{5/2}}$ and an arbitrary superposition state $\ket{\Phi_{n=20}} = \ket{n=20, l \geq 3, j, m = 5/2}$, are used to denote the computational $\ket{1}$ and $\ket{0}$ states for the $E_2$ gate respectively \cite{Yu_atom_flux_2017}. (b) Schematic of the system whereby an Rubidium-87 atom is held between two spherical capacitors (with capacitance $C$) connected to an LC resonator with inductance $L$. This atomic states are manipulated by an electrostatic field $E$ in the $z$-direction using parallel plates with imposed voltage difference $U$. The atom forms part of Rydberg quantum computer shown as an atom array (red dots in the purple block).}
    \label{fig:forster_atom}
\end{figure}

Rydberg quantum computers use Rydberg ensembles with three levels to perform computations:
the ground state $\ket{g}$ (mapped to $\ket{0}$), the hyperfine state $\ket{h}$ (mapped to $\ket{1}$), and the auxiliary excited state~$\ket{r}$. 
For Rydberg atom systems based on Rubidium-87, the hyperfine levels are chosen as $\ket{5S_{1/2}, F=1} = \ket{g}$ and $\ket{5S_{1/2}, F=2} = \ket{h}$~\cite{Levine_2019,Wu_2021}. The auxiliary state $\ket{r}$ corresponds to $\ket{70S_{1/2}}$ and is used to initiate the Rydberg blockade. Direct transitions from $\ket{0}$ to $\ket{r}$ require sub-300nm wavelengths, which are challenging to achieve. Instead, an intermediate auxiliary state $\ket{6P_{3/2}}$ is used to facilitate the transfer. The hyperfine transition occurs at 6.8 GHz, while the $\ket{0} \rightarrow \ket{6P_{3/2}}$ transition is driven by a 420nm laser. The final transfer to $\ket{r}$ is achieved using a 1013nm laser. A schematic of these energy levels in the computational basis is shown in Fig. \ref{fig:forster_atom}a.

The apparatus that connects the Rydberg atom system to the flux system follows the work of \cite{Yu_atom_flux_2017}. The study denotes a system in which a Rubidium-87 atom is placed between two spherical capacitors connected to an LC resonator in which an electric field runs in the $z$-direction. Two states are used in the study, a state $22D_{5/2}$ populated when the Rydberg qubit is in the computational $\ket{1}$ state and an arbitrary $\ket{n=20,l\geq3,j,m=5/2}$ superposition state is populated when the Rydberg qubit is in the $\ket{0}$ computational basis. The transition frequency between the two states is approximately 3.2GHz. This research extends \cite{Yu_atom_flux_2017} by connecting the Rubidium-87 atom in the system to an additional array of Rubidium-87 atoms. 

Bridging the atom array and the capacitor-confined atom requires specific state transitions. The $\ket{22D_{5/2}}$ state must transition to the computational $\ket{1}$ state in the hyperfine level $\ket{5S_{1/2}, F=2}$. Similarly, the $\ket{n=20, l\geq3, j, m=5/2}$ state (denoted as $\Phi_{n=20}$) must transition to the computational ground state $\ket{0} = \ket{5S_{1/2}, F=1}$. These transitions require optical wavelengths around 300nm, corresponding to THz frequencies. While theoretically feasible, further experimental studies should investigate the precise transition frequencies between the hyperfine levels of $\ket{5S_{1/2}}$ and the $\ket{22D_{5/2}}, \Phi_{n=20}$ states, as well as the transition efficiency.

For optimal system performance, atoms in the array must be sufficiently spaced during single-qubit operations to minimize unwanted interactions that could induce decoherence. This requires avoiding the dipolar exchange by ensuring the F\"{o}rster defect remains off-resonant, $\Delta_F \neq 0$. The interaction potential follows $V_{dd} = C_3/R^3$, making it crucial to maximize the interatomic distance $R$ during single-qubit operations.

Conversely, for multi-qubit gates, atoms must be close enough to enable exchange interactions. To implement the universal CZ gate, the Rydberg blockade must be activated, requiring a non-resonant F\"{o}rster defect, $\Delta_F > |\mu^2|/R^3$, where $\mu$ is the dipole transition moment for $\ket{5S_{1/2}} \rightarrow \ket{70S_{1/2}}$. The corresponding energy shift is given by $\Delta E_\pm = \pm C_6/R^6$.  

Using \textit{ARC} \cite{ibali__2017}, the characteristic radius was found to be $R_{vdW} = 3.5\mu$m (see \cite{juan_supp}). The dipole-dipole and van der Waals coefficients were calculated as $C_3 = 32.45$ GHz$\cdot\mu$m$^3$ and $C_6 = 801.98$ GHz$\cdot\mu$m$^6$, respectively. To minimize dipole-dipole interactions during single-qubit gates, an interatomic distance of $R \gtrsim 9\mu$m is required. For multi-qubit gates, $R = 3.5\mu$m results in an energy shift $\Delta E_\pm = 801.98$ GHz, enabling the Rydberg blockade effect.

In order to numerically simulate the $E_2$ gate and perform optimal control on the system, one needs to represent the master equation, or the equation denoting the dynamics of the system, in \textit{Lindbladian} form \cite{lindblad_1976}. The master equation is denoted in  (\ref{eq:lindbladian_form}). The first part of the equation is the Liouville-von Neumann equation \cite{phd_goerz} describing the pure state evolution of a system and $\{A_k\}$ are the \textit{Lindblad operators} representing the coupling of the system to the environment and thus the dissipative process \cite{joachain_1983,Distributed_Leung,lindblad_1976,nielsen_chuang_2021}. 
\begin{equation}\label{eq:lindbladian_form}
    \mathcal{L}[\rho] = [\mathcal{H}, \rho] + i\sum_k \gamma_k \bigg(A_k \rho A_k^\dag - \frac{1}{2} \Big\{ A_k^\dag A_k,\rho\Big \}\bigg),
\end{equation}
For Rydberg atom systems, the Hamiltonian is given in (\ref{eq:ryd_ham}). Here, $\Omega$ represents the rotation vector magnitude, $V_0$ the van der Waals interaction, and $|i - j|$ the interatomic distance. The detuning is denoted as $\varepsilon^{(i)}$, while $\epsilon^{(i)}(t)$ represents the energy shift of the Rydberg state $i$. The energy states are labeled by subscripts, with $e$ for the excited state and $g$ for the ground state.
\begin{equation}\label{eq:ryd_ham}
    \mathcal{H}_R= \frac{\Omega}{2}\sum_i \Big(\sigma_{eg}^{(i)}+\sigma^{(i)}_{ge}\Big) + V_0\sum_{i<j}\frac{\sigma_{ee}^{(i)}\sigma_{ee}^{(j)}}{|i-j|^6} - \sum_i \varepsilon^{(i)}(t)\sigma_{ee}^{(i)}.
\end{equation}
Assuming a local laser is used when addressing qubits in the quantum systems the local noise, the correlation between the phase noise experienced by atoms $i$ and $j$ at two different times is $\langle \varepsilon^{(i)}(t)\varepsilon^{(j)}(t') \rangle = (\Gamma/2)\delta_{ij}\delta(t-t')$, where $\Gamma$ is the dephasing rate, thus finding the local phase noise super-operator \cite{juan_supp,rydberg_noise, electromagnetic_gamma},
\begin{equation}\label{eq:rydberg_lindblad_1}
    \mathcal{L}[\hat{\rho}]^{l} = \Gamma \sum_{i}\bigg[\sigma_{ee}^{(i)}\hat{\rho}\sigma_{ee}^{(i)} - \frac{1}{2}\{\sigma_{ee}^{(i)}\sigma_{ee}^{(i)},\hat{\rho}\}\bigg].
\end{equation}

The final decoherence effect from the system comes from energy decay or the rate at which the population from an excited state $\ket{e}$ is transferred to a lower lying level $\ket{g}$ via photon emission into the vacuum. Spontaneous emission can be derived through Weisskopf-Wigner theory assuming complete coupling with closely spaced cavity modes with the emission spectrum centered at the atomic transition frequency. Weisskopf-Wigner theory also highlights that phenomenologically, the excited state is only capable of emitting to the ground state but not vice versa \cite{scully_zubairy_1997}. The final superoperator that describes the spontaneous decay from the state $\ket{e}$ reads,
\begin{equation}\label{eq:rydberg_lindblad_2}
    \mathcal{L}[\hat{\rho}]^{e} = \gamma_e\sum_i\Big[\sigma_{ge}^{(i)}\hat{\rho}\sigma_{eg}^{(i)} - \frac{1}{2}\{\sigma_{eg}^{(i)}\sigma_{eg}^{(i)},\hat{\rho}  \} \Big],
\end{equation}
where $\gamma_e$ is the spontaneous emission rate from the excited state. Using ARC \cite{ibali__2017}, the spontaneous decay time for the $\ket{e} = \ket{70S_{1/2}}$ is $375 \mu$s. The Lindbladians observed in (\ref{eq:rydberg_lindblad_1}) and (\ref{eq:rydberg_lindblad_2}) will be used to model the noise.

\subsection{Flux Qubit System}\label{sec:flux_system}

\begin{figure}[htbp]
\centerline{\includegraphics[width=0.5\textwidth]{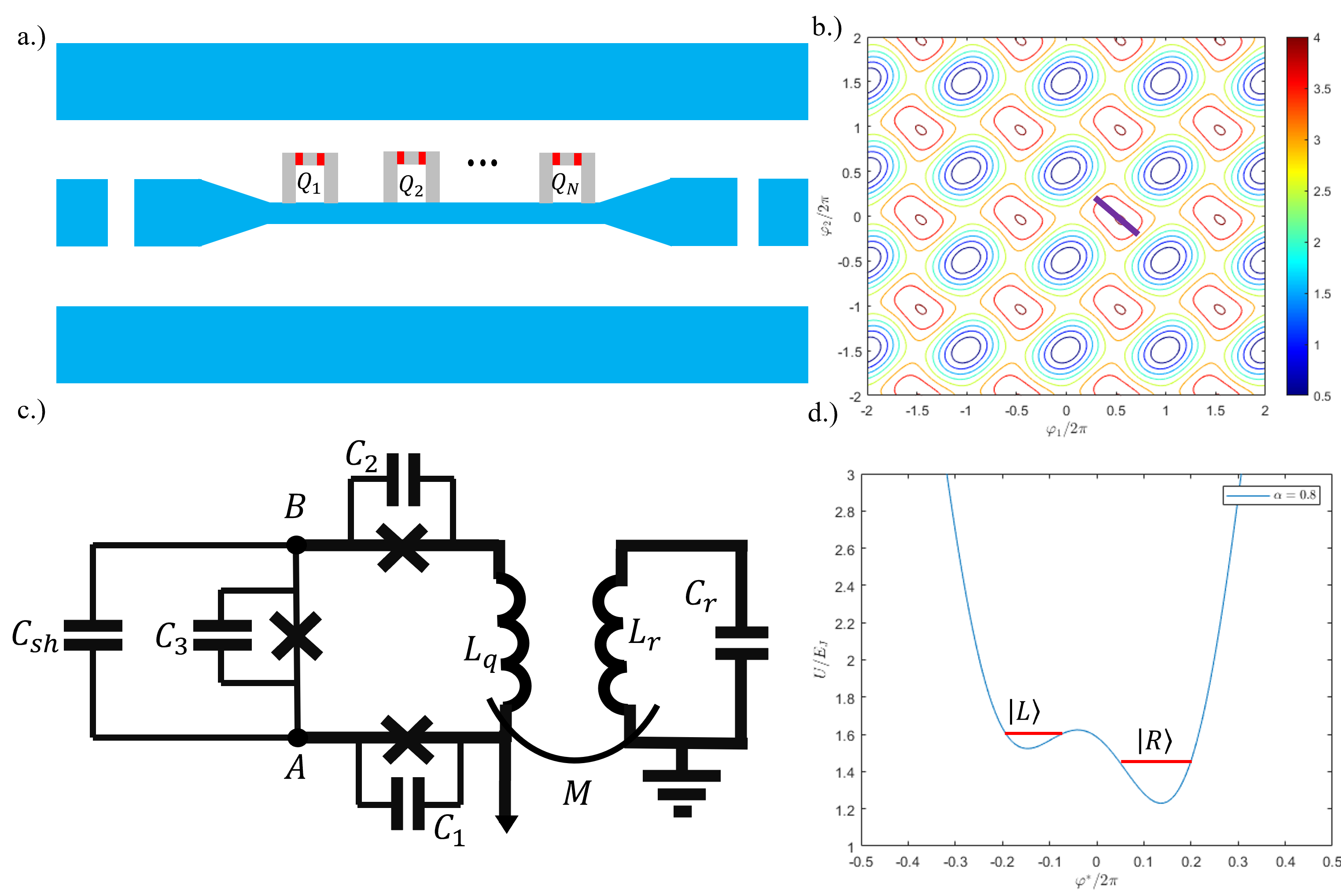}}
\caption{(a) Representation of Flux qubits (each qubit represented by $Q_i$) on a coplanar waveguide (in blue). (b) A contour plot of the potential energy $U/E_J$ at $f_\varepsilon = 0.53$ for $\alpha = 0.8$. The measured potential well is seen highlighted with the purple line. (c) A more specified outline of the C-shunt 3JJ flux qubit (circuit on the left) mutually coupled to a resonator, or in this case, the co-planar wave-guide (circuit on the right). (d) The potential function plotted against $\phi^* = \phi_1 = -\phi/2$ at $\alpha = 0.8$ and $f_\varepsilon = 0.53$ for the potential well seen in (c). The asymmetric bias gives rise to the $\ket{L}$ and $\ket{R}$ states.} 
    \label{fig:flux_system}
\end{figure}

The Josephson persistent-current qubit or the flux qubit is a type of superconducting qubit which consist of an inductance loop shorted by a Josephson junction \cite{pauw_flux}. The flux qubit's energy states are dependent on the magnetic flux measured in the inductance loop, which is dependent on the amount of Cooper pairs crossing the Josephson junction. The energetics of a single flux qubit system can be described by the Hamiltonian in (\ref{eq:single_JJ_Hamil}) where $Q_J$ is the electric charge on the capacitor, $C_J$ is the self capacitance, $I_q$ is the current circulating in the loop, $L$ is the inductance of the inductor, $\Phi$ is the applied magnetic field, $E_J$ is the energy from the Josephson junction and $\Phi_0$ is the magnetic flux quantum.

\begin{equation}\label{eq:single_JJ_Hamil}
    H = \frac{Q^2_J}{2C_J} - E_J\mathrm{cos}\Big(\frac{2\pi\Phi}{\Phi_0}\Big) + \frac{1}{2}LI_q^2.
\end{equation}
Following (\ref{eq:single_JJ_Hamil}), it is possible to see minimas form at $\Phi = \pm\Phi_0/2$. The two points are called the 'degeneracy points', in which the self-inductance of the Josephson junction is approximately equal to the inductance of the inductive loop $L_J \approx L$ \cite{Birenbaum2014TheCF}. These two values are associated with equal and opposite flux states formed by a persistent current $I_q$ in the loop. Because the magnetic field is seen to rotate clockwise or anti-clockwise around the loop the states are labeled left circle or $\ket{L}$ at $+\Phi_0/2$ and right circle or $\ket{R}$ at $-\Phi_0/2$ \cite{Birenbaum2014TheCF,pauw_flux}.

Maximising the trade-off between barrier height and the interaction between the states because the barrier height scales exponentially with $L/L_J$ where $L_J$ is the self-inductance of the Josephson junction. However, to form a double well potential form a single Josephson junction $L \approx L_J$. For this reason, it is best to use a three Josephson junction (3JJ) flux qubit which follows the ratio seen in (\ref{eq:alpha}), where $I_i$ is the critical currents of each junction, which follow the current-phase relation $I_i = I_0\mathrm{sin}\varphi_i$, where $\varphi_i$ are the gauge-invariant phase of each junction $i$ \cite{orlando_flux}. In this architecture, the ratio of junction sizes is important as opposed to the inductance \cite{Birenbaum2014TheCF}.
\begin{equation}\label{eq:alpha}
    \alpha = \frac{2I_3}{I_1+I_2},
\end{equation}

The ratio $\alpha$ and the magnetic frustration $f_\epsilon = \Phi/\Phi_0$ change the energy barrier and the anharmonicity of the qubit \cite{juan_supp}. This feeds into the two-level Hamiltonian of the 3JJ flux qubit system. In the two-level approximation the qubit Hamiltonian is described in terms of the magnetic energy $E_{LR}$ and the tunnel coupling $\Delta/2$ that creates an anti-crossing \cite{pauw_flux}. In the persistent-current basis ($\ket{L},\ket{R}$) the qubit Hamiltonian is,
\begin{equation}\label{eq:easy_two_level}
    H = -\frac{1}{2}(\varepsilon\sigma_z + \Delta \sigma_x),
\end{equation}
where $\varepsilon = 2E_{R} = 2I_p(f_{\varepsilon}-1/2)\Phi_0$, $I_p$ is the persistent current, and both $\sigma_x$ and $\sigma_z$ are the Pauli matrices \cite{pauw_flux}. The corresponding eigenenergies of the diagonalized Hamiltonian are $E_0,E_1 = \mp\frac{1}{2}\sqrt{\varepsilon^2+\Delta^2}$ where the energy level splitting in the computational basis, $\{0,1\}$, is $E_{01} = E_1 - E_0 =\sqrt{\varepsilon^2+\Delta^2}$.

In order to reduce noise, flux qubits coupled together via transmission lines or resonators, tend to possess an additional capacitor. This is called \textit{C-shunted} flux qubits \cite{Birenbaum2014TheCF,Yan_2016}. This is due to the reduced noise susceptibility that is obtained when introducing an extra capacitor. The modelled circuit can be seen in Fig. \ref{fig:flux_system}(c) coupled via a resonator to the co-planar waveguide (CPW). Fig. \ref{fig:flux_system}(a) shows the flux qubits geometrically on a CPW. 

The two-level-system Hamiltonian for an individual C-shunt flux qubit near flux-degeneracy and coupled to a CPW resonator is,
\begin{equation}\label{eq:approximated_model}
    H = \frac{\hbar}{2}[\varepsilon\sigma_z + \Delta\sigma_x] + \hbar \omega(\hat{a}^\dag \hat{a} + \frac{1}{2}) + \hbar g\sigma_y(\hat{a}^\dag + \hat{a}),
\end{equation}
where the first part of the equation follows \ref{eq:easy_two_level} and $\hat{a}^\dag (\hat{a})$ is the raising (lowering) operator for photons. This Hamiltonian can be intuitively broken down into three different terms. The three terms are respectively, the qubit, resonator and qubit-resonator Hamiltonians \cite{Yamamoto_2014,Yan_2016}. The coupling strength between the resonator and qubit are dependent on $M$, $L_r$, $C_r$ and $L_q$ \cite{nonlinear_koun,Harris_2007}.

The C-shunt flux qubit has three features that differentiate it from the normal 3JJ flux qubit. It has a lower critical current, $I_c$, typically $\alpha < 0.5$ and the additional shunting junction has a capacitance $C_{sh} = \zeta C$, where $C$ is the capacitance of the smallest junction and $\zeta \gg 1$. The 3JJ capacitively-shunted flux qubit can be described by a Hamiltonian with a kinetic and potential part,
\begin{flalign}\label{eq:kinetic_energy_flux}
    H & = T + U, \\
    T & = \frac{1}{2}(\boldsymbol{Q}+\boldsymbol{q})^\intercal \mathbf{C}^{-1}(\boldsymbol{Q}+\boldsymbol{q}), \\
    U & = E_J\{ 2 + \alpha - \mathrm{cos}\varphi_1 - \mathrm{cos}\varphi_2-\alpha\mathrm{cos}(2\pi f_\varepsilon + \varphi_1 - \varphi_2) \}. 
\end{flalign}
The charges $\boldsymbol{Q}$ and $\boldsymbol{q}$ are the charges induced charges on the islands, where the islands refer to the two nodes $A$ and $B$ in Fig. \ref{fig:flux_system}(c) \cite{Yan_2016}. The matrices can be rewritten to,
\begin{equation}
    \boldsymbol{Q} = -2e\begin{bmatrix}
        \frac{\partial}{\partial \varphi_1} \\
        \frac{\partial}{\partial \varphi_2}
    \end{bmatrix},
    \quad
    \boldsymbol{q} = \begin{bmatrix}
        q_A \\
        q_B
    \end{bmatrix},
\end{equation}
\begin{equation}
    \boldsymbol{C} = C\begin{bmatrix}
        \zeta + 1 + \alpha & -(\zeta + \alpha) \\
        -(\zeta + \alpha) & \zeta + 1 + \alpha
    \end{bmatrix}.
\end{equation}
The system is less sensitive to charge fluctuations due to the shunt capacitor reducing the effective charging energy. Choosing $\varphi_+ = (\varphi_1 + \varphi_2)/2$ and $\varphi_- = (\varphi_1 - \varphi_2)/2$ using the Cooper-pair number operators $\hat{n}_{\sigma} = -i\partial/\partial\varphi_i$ where $i = \{-,+\}$, the reduced Hamiltonian becomes,
\begin{equation}
\begin{split}
    H &= \frac{1}{2}E_{C,+}\hat{n}^2_+ + \frac{1}{2}E_{C,-}\hat{n}^2_- \\
    & + E_J\{2+\alpha-2\mathrm{cos}\varphi_-\mathrm{cos}\varphi_+ -\alpha\mathrm{cos}(2\pi f_\varepsilon + 2\alpha_-)\},
\end{split}
\end{equation}
where $E_{C,+} = 2e^2/C$ and $E_{C,-} = (\zeta+\alpha+1)e^2/C$ are the effective charging energy for the $+$-mode and $-$-mode respectively. Ideally, the $+$-mode can be omitted since $\zeta \gg 1$. Thus, the simplified Hamiltonian is \cite{Yan_2016},
\begin{equation}\label{eq:flux_hami}
    H_m = \frac{1}{2}E_{C,-}\hat{n}^2_- + E_J\{-2\mathrm{cos}\varphi_- -\alpha\mathrm{cos}(2\pi f_\varepsilon + 2\alpha_-)\}.
\end{equation}

The flux qubit noise can be placed in two categories. \textit{Charge fluctuations} are the noise generated from effects on the circuit such as decoherence from qubit coupling effects \cite{pauw_flux} and \textit{flux fluctuations} are the non-circuit related noise such as fluctuations in the magnetic field or the Josephson junction itself. Performing a first-order perturbation on the derived Hamiltonian in (\ref{eq:flux_hami}) can be used to analytically model the flux fluctuations.
\begin{equation}
    \begin{split}
        \delta\mathcal{H} &= \delta U \\
        &\approx -2\pi\alpha E_J(\mathrm{sin}\phi (1-2\varphi_z^2(\hat{a} + \hat{a}^\dag))\delta f,
    \end{split}
\end{equation}
Following (\ref{eq:flux_hami}), $\varphi_z = (E_{C,-}/4E_{J,-})^{1/4}$, $E_{C,-}$ is the effective charging energy for the negative minimum, $E_{J,-}$ is the effective Josephson energy in the negative minimum and $\phi = \phi(f_\varepsilon) = 2\pi f_\varepsilon + 2\varphi_-^*(f_\varepsilon)$.

Fluctuations near the frequency $\Omega$ are considered because they have the possibility of inducing transitions between the $\ket{L}$ and $\ket{R}$ states. Thus, in the case $\delta f \propto \mathrm{cos}(\Omega t)$, the system simplifies to two levels,
\begin{equation} \label{eq: basic_hamil_flux}
    \mathcal{H}= \frac{1}{2}(\hbar\omega_q\sigma_z+I_m \Phi_0\delta f\sigma_x)
\end{equation}
where $\omega_q = \omega_q(f_\varepsilon) = \Omega$ and $I_m$, the $f_\varepsilon$-dependent current difference between the parametrized circulating-current states, is defined as $I_m = I_m(f_\varepsilon) = -8\pi\alpha\sigma_z\mathrm{cos}\phi E_J/\Phi_0$.

Investigating the charge fluctuations ($\delta q_a$,$\delta q_b$) requires invoking a perturbation via the kinetic energy $T$. Perturbing the kinetic energy given by \eqref{eq:kinetic_energy_flux} yields,
\begin{equation}
    \begin{split}
        \delta T &= \delta \boldsymbol{q}^\intercal\boldsymbol{C}^{-1}\boldsymbol{Q} \\ 
        &= -\frac{e}{C(2\zeta+2\alpha+1)} [-in_z(\hat{a}-\hat{a}^\dag)](\delta q_A - \delta q_B),
    \end{split}
\end{equation}
where $n_z = (E_{J,-}/4E_{C,-})^{1/4}$ is the quantum ground-state uncertainty in Cooper-pair number and $\zeta$ is called the shunt factor which relates the capacitance of the smaller junction ($C_3$ seen in Fig. \ref{fig:flux_system}(c)) to the capacitance of the shunt capacitor through $C_{sh}=\zeta C_3$ \cite{Yan_2016}. This factor is considered to satisfy $\zeta \gg 1$. The charge noise is orthogonal to the flux noise, thus it couples to the Hamiltonian in the $\sigma_y$ basis. The perturbation depends only on the differential mode of the induced charges between the two islands (described by nodes $A$ and $B$ seen in Fig. \ref{fig:flux_system}(c)) \cite{Yan_2016}. Thus, placing the charge fluctuation back into \eqref{eq: basic_hamil_flux}, gives,
\begin{equation}\label{eq:flux_fluctuation_final}
    \mathcal{H} = \frac{1}{2}(\hbar \omega_q \sigma_z + I_m \Phi_0 \delta f \sigma_x + n_z E_{C,-}\delta n_- \sigma_y),
\end{equation}
where $\delta n_- = (\delta q_A - \delta q_B)/(-e)$ is the differential electron number fluctuation \cite{Yan_2016}. 

The importance of the C-shunted flux design is observed in the noise sensitivity. Because of this design, the charge noise sensitivity diminishes with larger $\zeta$ due to the relation, $n_z E_{C,-} \propto E^{3/4}_{C,-} \propto \zeta^{-3/4}$. Each one of these fluctuation terms can be applied as a Lindbladian to the Hamiltonian seen in \eqref{eq:approximated_model}. The Lindbladians modeling the noise due to the fluctuations in flux and charge noise in the $\sigma_x,\sigma_y,\sigma_z$ Pauli bases, are thus \cite{Yan_2016},
\begin{equation}\label{eq:flux_linblad_1}
    \mathcal{L}[\hat{\rho}]_z = \frac{1}{2}\omega_q\Big[\sigma_z\hat{\rho}\sigma_z^\dag - \frac{1}{2}\{\sigma_z^\dag\sigma_z, \hat{\rho}\} \Big],
\end{equation}
\begin{equation}\label{eq:flux_linblad_2}
    \mathcal{L}[\hat{\rho}]_x = \frac{1}{2\hbar}I_m\Phi_0\delta f\Big[\sigma_x\hat{\rho}\sigma_x^\dag - \frac{1}{2}\{\sigma_x^\dag\sigma_x, \hat{\rho}\} \Big],
\end{equation}
\begin{equation}\label{eq:flux_linblad_3}
    \mathcal{L}[\hat{\rho}]_y = \frac{1}{2\hbar}n_z E_{C,-}\delta n_-\Big[\sigma_y\hat{\rho}\sigma_y^\dag - \frac{1}{2}\{\sigma_y^\dag\sigma_y, \hat{\rho}\} \Big].
\end{equation}

The final non-circuit noise source comes from the CPW coupled to the flux qubit. The CPW has been simplified to operate similarly to a resonant cavity. Due to this, the system encounters noise in the form of the Purcell effect. The Purcell effect is the enhancement of a quantum system's spontaneous emission rate due to its environment. In this case, the CPW radiates a wave which has been reflected from the environment which consequently excites the resonance out of phase. This is difficult to analytically model due to the vast sources of electromagnetic waves possible, thus, using experimental results~\cite{Yamamoto_2014}, the rate at which the resonator encounters spontaneous emission is modeled by,
\begin{equation}
    \frac{1}{\tau_{photon}} = \frac{\sqrt{A \mathrm{ln} 2}}{\hbar} \Big| \frac{\delta \omega_{LR}}{\delta f_\varepsilon} \Big|,
\end{equation}
where $\omega_{LR}$ is the transition frequency between the states $\ket{L}$ and $\ket{R}$, and $A$ is a fit parameter. According to \cite{Yamamoto_2014}, $1/\tau_{photon}$ was measured to be around 9.19 MHz for $0.5<f_\varepsilon<0.55$. The Lindbladian can, thus, be described as,
\begin{equation}\label{eq:flux_linblad_4}
    \mathcal{L}[\hat{\rho}]_{purcell} = \frac{1}{\tau_{photon}}\Big[\sigma_y\hat{\rho}\sigma_y^\dag - \frac{1}{2}\{\sigma_y^\dag\sigma_y, \hat{\rho}\} \Big].
\end{equation}
The Lindbladians observed in \eqref{eq:flux_linblad_1}, \eqref{eq:flux_linblad_2}, \eqref{eq:flux_linblad_3} and \eqref{eq:flux_linblad_4} will be used to model the noise.

To realize the full potential of a C-shunt flux qubit, one would need to choose $\alpha<0.5$ because it reduces the circuit sensitivity to charge noise and stray fields \cite{Birenbaum2014TheCF,UltrastrongCapacitiveCoupling}. However, in this regime, the qubit resembles a phase qubit where the potential well becomes singular and thus loses anharmonicity between two energy states. In order to realize the hybrid $E_2$ gate, each flux qubit requires two potential wells with states $\ket{L}$ and $\ket{R}$ in each \cite{Xiang_2013}. Additionally, the potential wells must have differing depths as seen on Fig. \ref{fig:flux_system}(b) and \ref{fig:flux_system}(d). Maintaining the anharmonicity between the two levels allows there to be a very strong and a very weak coupling to a resonator depending on the energy level. It is important that the flux qubit system only couples to the Rydberg system when the resonator is in the $\ket{1}$ state, or when the system needs to communicate. Hence, there is a trade-off between noise and the coupling to the hybrid system \cite{Xiang_2013}. Thus, the chosen values for the flux system are $\alpha = 0.8$ and $f_\varepsilon = 0.53$ as seen in Fig. \ref{fig:flux_system}(b) and \ref{fig:flux_system}(d).

\subsection{Indirect Coupling \& Non-Local GHZ State}
This study focuses on a simplified model where a chain of flux qubits and a chain of Rydberg atoms are coupled to the same LC resonator, inspired by \cite{Yu_resonator_atom_2016}. While similar to the Hamiltonians in Sec. \ref{sec:rydberg_atom_system} and \ref{sec:flux_system}, a simpler model is used since the $E_2$ gate operates on a much shorter timescale than previously modeled effects \cite{Yu_atom_flux_2017}. This extensively studied system \cite{Yu_atom_flux_2017} will be used to establish the $E_2$ gate.

\begin{figure}[htbp]
\centerline{\includegraphics[width=0.5\textwidth]{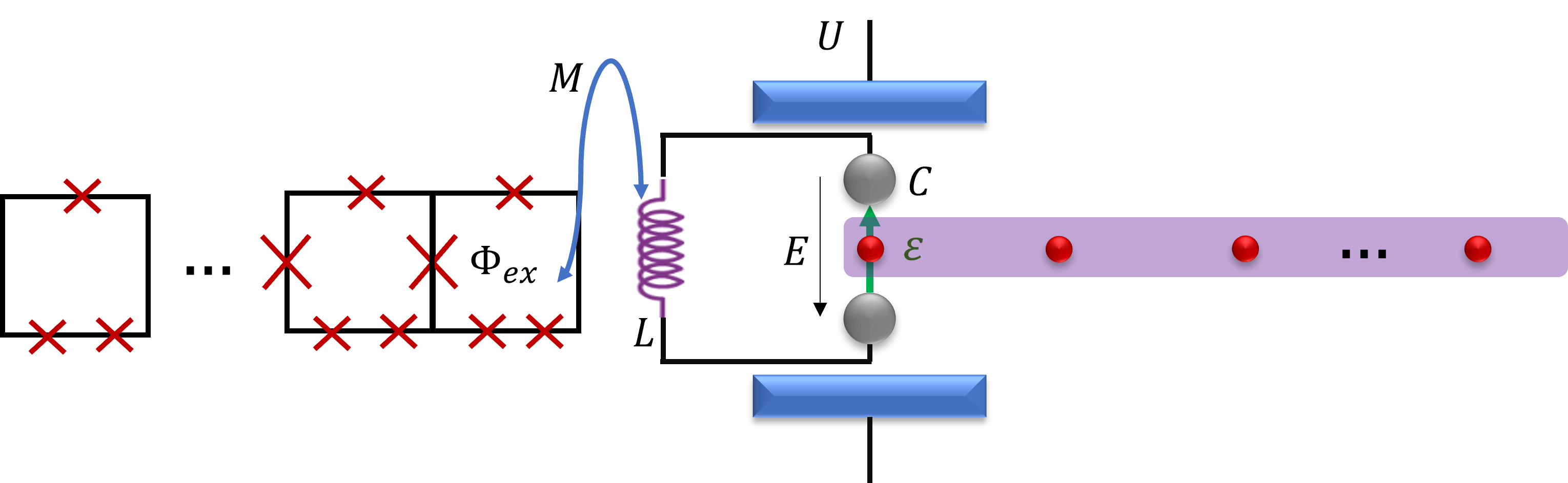}}
    \caption{Hybrid system composed of a chain of three-JJ flux qubits, an LC resonator and an atom. The flux qubits have a tunable coupling mitigated by Josephson coupling while the Rydberg atom is Rydberg coupled to a chain of Rydberg atoms. The flux qubit, biased by an external flux $\Phi_{ex}$ is inductively coupled to the resonator with a mutual inductance $M$. The atom is placed in the middle point between two spheres with an oscillating intra-resonator electric field with amplitude $\varepsilon$ in the $z$-axis. An additional electrostatic field $E$ runs across the Rydberg qubit in the $z$-axis generated by the parallel-plate capacitors. This model is inspired by \cite{Yu_resonator_atom_2016}.} 
    \label{fig:hybrid_model}
\end{figure}

A diagram of the model is shown in Fig. \ref{fig:hybrid_model}. The mediator is a superconducting LC resonator at mK temperatures with a characteristic frequency $\omega_0 = 1/\sqrt{LC}$, set to $2\pi \times 20$ GHz. The capacitor, composed of identical spherical spheres, has capacitance $C$ and inductance $L$. The numerical values for the parameters can be seen in \cite{juan_supp}.
\begin{equation}
    H_{LC} = \frac{\phi^2}{2L} + \frac{q^2}{2C} = \hbar \omega_0 (b^\dag b + 1/2),
\end{equation}
where $b^\dag$ and $b$ are the creation and annihilation operators, the magnetic flux $\phi = \sqrt{\hbar/2C\omega_0}(b^\dag + b)$ and $q = i \sqrt{C\hbar\omega_0 /2}(b^\dag - b)$ is the charge operator \cite{Yu_charge_qubit_2016,Yu_atom_flux_2017}. The Rydberg atom is placed in the midpoint between the two spherical capacitors which couples to the local electric field in the $z$-direction resulting in the atom-resonator interaction operator \cite{Yu_resonator_atom_2016,Yu_atom_flux_2017},
\begin{equation}
    V_a = -i D \varepsilon(b^\dag - b).
\end{equation}
$D$ is the atomic dipole moment and $\varepsilon$ is the amplitude of the oscillating electric field inside the resonator. It is assumed that the inhomogeneity of $\varepsilon$ within the atomic wave-pocket is $|\frac{1}{\varepsilon}\frac{\partial \varepsilon}{\partial \alpha}r| < 10^{-3}$ where $\alpha$ are the different dimensions $\alpha = x,y,z$ and $r$ is the radius of the Rydberg state, consequently affecting the atom-resonator coupling \cite{Yu_atom_flux_2017}. Furthermore, the inhomegeneity of the external electric field, $E$, caused by the screening effect of the spheres is also negligible.  

The Rydberg atom coupled to the resonator starts at the excited state $\ket{e}$ on $22^2D_{5/2}(m=5/2)$. When the electric field $E$ is around 500 V/cm, the excited state interacts with states in the set $\ket{n=20,l\geq3,j=l \pm \frac{1}{2}, m =\frac{5}{2}}$. In this electric field regime, there exists an excited state $\ket{e}$, a ground state $\ket{g}$ and an auxiliary state $\ket{u}$. Through adiabatic tuning, one can vary through the states $\ket{\mu = e,g,u}$. The Hamiltonian of the atom is given by,
\begin{equation}
\begin{split}
    H_a &= \sum_{\mu = e,g,u} \hbar \omega_\mu \ket{\mu}\bra{\mu}\\ 
    &+  \frac{\hbar\Omega}{2}(\ket{e}\bra{g} + \ket{g}\bra{e}) + \frac{\hbar\Omega^{'}}{2}(\ket{e}\bra{u} + \ket{u}\bra{e}),
\end{split}
\end{equation}
where $\omega_{\mu=e,g,u}$ are the electric-field-dependent energies of the atomic states \cite{Yu_atom_flux_2017}. The atom-resonator interaction is expressed as,
\begin{equation}
    V_a = \frac{\hbar g_a}{2}(b^\dag\ket{g}\bra{e} + \ket{e}\bra{g}b) + \frac{\hbar g^{'}_a}{2}(b^\dag\ket{u}\bra{e} + \ket{e}\bra{u}b),
\end{equation}
with coupling strengths $g_a = |\bra{e}D\ket{g}|\varepsilon/\hbar$ and $g_a^{'} = |\bra{e}D\ket{u}|\varepsilon/\hbar$ \cite{Yu_resonator_atom_2016, Yu_atom_flux_2017}.  

The three-JJ flux qubit is coupled to the LC resonator through means of mutual induction with inductance $M$. The flux qubit is biased by an external magnetic flux $\Phi$ to tune the frequency spacing $\varepsilon = 2I_p\Phi_0\gamma_q/\hbar$ with phase bias parameter $\gamma_q = \Phi/\Phi_0 -1/2$ as mentioned in Section \ref{sec:flux_system} \cite{Chiorescu_2003,Yu_atom_flux_2017}. The qubit is tuned between the states $\ket{R}$ and $\ket{L}$. Thus, the Hamiltonian for the flux-qubit is similar to \eqref{eq:easy_two_level} \cite{Yu_charge_qubit_2016,Yu_atom_flux_2017},
\begin{equation}
    H_f = -\frac{\hbar \varepsilon}{2}\sigma_{f,z} - \frac{\hbar \Delta}{2} \sigma_{f,x},
\end{equation}
but with the definitions of the Pauli matrices being, $\sigma_{f,z} = \ket{L}\bra{L} - \ket{R}\bra{R}$, $\sigma_{f,x} = \sigma^\dag_{f,-}+\sigma_{f,-}$ and $\sigma_{f,-} = \ket{R}\bra{L}$. The tunnel splitting is denoted as $\Delta$  with the flux qubit-resonator interaction potential given by,
\begin{equation}
    V_f = -\hbar g_f(b^\dag+b)\sigma_{f,z},
\end{equation}
where $g_f$ is the coupling strength \cite{Yu_atom_flux_2017}. The final system Hamiltonian is thus,
\begin{equation}\label{eq:final_hamil_hybrid}
    H = H_{LC} + H_a + H_f + V_a + V_f.
\end{equation}

The preparation of the non-local GHZ state is detailed in depth in \cite{Yu_atom_flux_2017} with a reported 8ns time with a fidelity of 0.977. The steps from \cite{Yu_atom_flux_2017} were numerically simulated in the supplementary material \cite{juan_supp}. Using QuTiP \cite{qutip_package}, the numerical simulation was shown to achieve a 0.93 fidelity in 17ns which is roughly in accordance with the scales in \cite{Yu_atom_flux_2017}. 

The coupling system possesses four relevant sources of noise. The flux qubit has been simplified to contain two noise sources, qubit dephasing, given by a dephasing rate, as well as a relaxation loss of a qubit decohering from $\ket{R}$ state to a $\ket{L}$ state, given by a relaxation rate \cite{Yu_atom_flux_2017}. The dephasing rate and relaxation rate are denoted by $\gamma_{relax}$ and $\gamma_\phi$, respectively. The main noise source observed in the hybrid system is the spontaneous decay rate. The decay is observed for the states $\mu = e,u$ to the ground state $\ket{g}$. The spontaneous emission rate is assumed to be equal for all $\mu$ because the Rydberg state is close to the capacitor surface which induces extra noise from stray fields and thus drastically reduces the lifetime of the states \cite{Yu_atom_flux_2017}. Finally, the noise of the resonator is characterized by $\kappa = \omega_0/Q$ (assuming a Q-factor value $Q = \mathrm{10^5}$) as the loss rate due to the Purcell effect. Thus, the coupling system possesses the following Lindbladians \cite{Yu_atom_flux_2017},
\begin{equation}
    \mathcal{L}^f_{relax} = \gamma_{relax}\Big[\sigma_{f,-}\rho\sigma_{f,-}^\dag - \frac{1}{2}\{\sigma_{f,-}^\dag\sigma_{f,-},\rho\}\Big]
\end{equation}
\begin{equation}
    \mathcal{L}^f_{dephase} = \frac{\gamma_\phi}{2}\Big[\sigma_{f,z}\rho\sigma_{f,z}^\dag - \frac{1}{2}\{\sigma_{f,z}^\dag\sigma_{f,z},\rho\}\Big]
\end{equation}
\begin{equation}
    \mathcal{L}^r = \kappa\Big[b\rho b^\dag - \frac{1}{2}\{b^\dag b,\rho\}\Big]
\end{equation}
\begin{equation}
    \mathcal{L}^a = \gamma_{e} \sum_{\mu = e,u}\Big[\sigma_{\mu g}\rho\sigma_{\mu g}^\dag - \frac{1}{2}\{\sigma_{\mu g}^\dag\sigma_{\mu g},\rho\}\Big].
\end{equation}
The final fidelity with noise is very similar to the fidelity without noise at the same time scale. This is due to the fact that the decoherence timescale is a factor of $\mathrm{10^3}$ larger than the creation of the $E_2$ gate. 

\section{GRAPE Optimized Distributed Phase Estimation}
This section describes uses Quantum Optimal Control Theory, and specifically the GRAPE algorithm to optimize the qubit gates in the non-local phase estimation algorithm.

\subsection{The GRAPE Algorithm}
Simulating quantum algorithms with many gates on a noisy open system is infeasible due to rapid decoherence. While one could analytically solve the master equation and adjust gate parameters accordingly, this requires solving the Schr\"{o}dinger equation at each step, which becomes impractical for large qubit systems.

In the NISQ era, optimizing qubit gates through optimal control has become a key focus \cite{Koch_2022}. Quantum Optimal Control Theory (QOCT) tailors external electromagnetic fields to efficiently steer quantum dynamics. Among various techniques, Gradient Ascent Pulse Engineering (GRAPE) \cite{KHANEJA2005296} is particularly effective due to its use of analytical gradient expressions, enabling rapid convergence in parameter optimization \cite{chen2022iterative,KHANEJA2005296}. 
To apply GRAPE, it is essential to first characterize the system. The Hamiltonian consists of a drift term and a control term, as shown in \eqref{eq:control_drift_ham}. The control term can be further expressed as a sum of time-dependent control coefficients $c_j(t)$ acting on a time-independent Hamiltonian.
\begin{equation}\label{eq:control_drift_ham}
    H(t) = H_d(t)+ H_c(t) = H_d(t) + \sum_jc_j(t)H_j
\end{equation}
To model the system into an open system, one must first utilize the density operator $\rho (t)$ and characterize the equation of motion as a Liouville-von Neumann equation,
\begin{equation} \label{eq:density_init}
    \dot{\rho}(t) = -\frac{i}{\hbar} \bigg [ \Big(\mathcal{H}_0 + \sum^m_{k=1}u_k(t)\mathcal{H}_k\Big), \rho(t)\bigg],
\end{equation}
where $\mathcal{H}_0$ is the drift Hamiltonian, $\mathcal{H}_k$ are the radio-frequency Hamiltonians that correspond to the available control fields and $u(t) = (u_1(t), u_2(t),\cdots,u_m(t))$ is the control vector or a set of vector amplitudes that can be changed \cite{KHANEJA2005296}. The goal is to find the optimal amplitudes $u_k(t)$ of the RF fields capable of steering a given density operator in initial state $\rho(0) = \rho_0$ in a total time $T$ to a density operator $\rho(T)$ with maximum overlap to some desired target $C$ measured by the standard inner product,
\begin{equation}\label{eq:trace_func}
    \braket{C|\rho(T)} = \mathrm{Tr}\{ C^\dag \rho(T) \}.
\end{equation}
In the case of open quantum systems, the dynamics is described by the Markovian master equation seen in \eqref{eq:lindbladian_form}. The formal solution to such equation is \cite{Boutin_2017}, 
\begin{equation}
    \rho(t) = \mathrm{exp}\Big\{\int^t_0\mathcal{L}(t^{'})dt^{'}\Big\}\rho(0).
\end{equation}
To iterate through time, the transfer time $T$ is discretiszd in $N$ equal steps of duration $\Delta t = T/N$. It is assumed that at each time step, the control amplitudes $u_k(t)$ are constant. This means that at an arbitrary time step $l$, the amplitude $u_k(t)$ of the $k$th control Hamiltonian is given by $u_k(l)$. Using the Lindbladian formulation described Eq. \eqref{eq:lindbladian_form}, the master equation is,
\begin{equation}
    \dot{\rho} = -i[H,\rho] + \mathcal{L}[\rho],    
\end{equation}
where $\mathcal{L}$ acts as an operator on a density matrix, $\rho$. Using this definition means that one can define a discretized unitary propagator during a time step $l$,
\begin{equation}\label{eq:unitary_grape}
    U_l \boldsymbol{\cdot} = \mathrm{exp} \bigg\{-i\Delta t([H_l, \boldsymbol{\cdot}] + \mathcal{L}[\boldsymbol{\cdot}] \bigg \}.
\end{equation}
A performance index is required for the algorithm to optimize a gate, which can be expressed as the overlap between a final state $\rho(T)$ and the target state, $C$. The performance index is of the form,
\begin{equation}
    \Phi_0 = \mathrm{Tr}\Big\{C U_N\cdots U_1\rho(0)\Big\}.
\end{equation}
The derivative of the performance index takes the form \cite{Boutin_2017},
\begin{equation}
    \frac{\partial \Phi_0}{\partial u_k(l)} = \mathrm{Tr} \bigg\{\lambda_l(C)\frac{\partial U_l}{\partial u_k(l)}\rho_{l-1} \bigg\},
\end{equation}
where,
\begin{equation}
    \rho_l = U_l \cdots U_1 \rho(0),
\end{equation}
is the forward in time evolved density matrix and,
\begin{equation}
    \lambda_j(C) = U_{l+1}^\dag\cdots U_N^\dag C,
\end{equation}
is the backwards in time evolution from the final target state.The first order derivative of the $l^{th}$ time-evolution operator in terms of $\Delta t$ is~\cite{Boutin_2017,Abdelhafez_2019},
\begin{equation}
    \frac{\partial U_l \cdot}{\partial u_l(l)} \approx -i\Delta t [H_k, (U_l \boldsymbol{\cdot})].
\end{equation}
The derivative of the performance index is thus,
\begin{equation}\label{eq:GRAPE_lindblad}
    \frac{\partial\Phi_0}{\partial u_k(l)} = -i \Delta t \mathrm{Tr}\{\lambda_l(C)[H_k,\rho_l]\}.
\end{equation}
The result seen in Eq. \eqref{eq:GRAPE_lindblad} is the core of the GRAPE algorithm ~\cite{Abdelhafez_2019,Boutin_2017}. 

The efficiency of the algorithm is dependent on maximizing the performance gradient $\partial\Phi_0/\partial u_k(l)$. The performance change of the control amplitude is acquired by allowing \cite{chen2022iterative}, 
\begin{equation} \label{eq:u_k_def}
    u_k(l) \rightarrow u_k(l) + \epsilon\ \frac{\delta \Phi_0}{\delta u_k(l)},
\end{equation}
so that the performance function $\Phi_0$ increases when $\epsilon$ is a small enough step size. Optimizing the GRAPE algorithm for qubit control requires redefining the performance function $\Phi_0$. In qubit optimal control, this function, given by \eqref{eq:trace_func}, is also known as the fitness function, $f$~\cite{Lu_2017}, where $\varepsilon = 1 - f$. As fidelity improves, modifications decrease. For a full evolution, there are $N \times M$ variables, with $M$ representing the number of control Hamiltonians. The general GRAPE algorithm is outlined in Alg. \ref{alg:grape_basic}.

\begin{algorithm}
\caption{The general GRAPE algorithm.}\label{alg:grape_basic}
\begin{algorithmic}
\Require Get minimum fidelity error $x_{min}$
\Require Guess initial controls $u_k(l)$
\State $\rho \gets \rho_0$
\State $done \gets \mathrm{False}$
\While{not $done$}
    \State $l \gets 0$
    \While{$l \leq N$}
        \State $Iterator_{f} \gets 0$
        \While{$Iterator_{f} \leq l$}
            \State $\rho \gets U_{Iterator_{f}} \rho$ 
            \State $Iterator_{f} \gets Iterator_{f} + 1$
        \EndWhile
        \State $\lambda \gets C$
        \State $Iterator_{b} \gets N$
        \While{$Iterator_{b} > l+1$}
            \State $\lambda \gets U^\dag_{Iterator_{b}}$
            \State $Iterator_{b} \gets Iterator_{b} + 1$
        \EndWhile
        \State $l \gets l + 1$
    \EndWhile
    \State Evaluate $\partial\Phi_0/\partial u_k(l)$ (Eq. \eqref{eq:GRAPE_lindblad}), $\varepsilon=1-f$ and update $m \times N$ control amplitudes $u_k(l)$ according to Eq. \eqref{eq:u_k_def}
    \If{ $\varepsilon \leq x_{min}$ }
        \State $done \gets \mathrm{True}$
    \EndIf
\EndWhile

\end{algorithmic}
\end{algorithm}

A key challenge for GRAPE and other optimization algorithms is their tendency to get stuck in local minima. For systems with many control fields $\mathcal{H}_k$, this is manageable by defining fidelity with sufficient precision. However, for systems lacking necessary control fields, the maximum achievable fidelity is unknown, creating a trade-off between iteration count and the risk of overstepping a minimum.

To mitigate this, the local problem landscape can be mapped to a parabola, enabling optimal step sizes without guesswork. GRAPE employs the Broyden-Fletcher-Goldfarb-Shanno (BFGS) algorithm, a quasi-Newtonian gradient descent method that approximates Hessians efficiently \cite{LBFGS,compareBFGS,Dalgaard_2020,BFGS_practical,wolfe_condition}.

The controls used for the GRAPE algorithm are the Pauli matrices $\sigma_x, \sigma_y$ and $\sigma_z$. For the flux system, the drift Hamiltonian will be \eqref{eq:approximated_model} with the superoperators \eqref{eq:flux_linblad_1} - \eqref{eq:flux_linblad_4}. For the Rydberg atom system, the drift Hamiltonian will use \eqref{eq:ryd_ham} with the superoperators \eqref{eq:rydberg_lindblad_1} and \eqref{eq:rydberg_lindblad_2}. These master equations will be fed into the GRAPE algorithm that already form part of the QuTiP optimal control package \cite{juan_supp, qutip_package}. The GRAPE algorithm is capable of achieving correction steps in the order of 10-100ns which means that time steps are too large for the preparation of the $E_2$ gate. However, as was aforementioned, the decoherence timescale is much larger than the timescale of the gate creation, thus GRAPE is not needed.   

\begin{figure}[t] 
\centering
    \centerline{\includegraphics[width=0.5\textwidth]{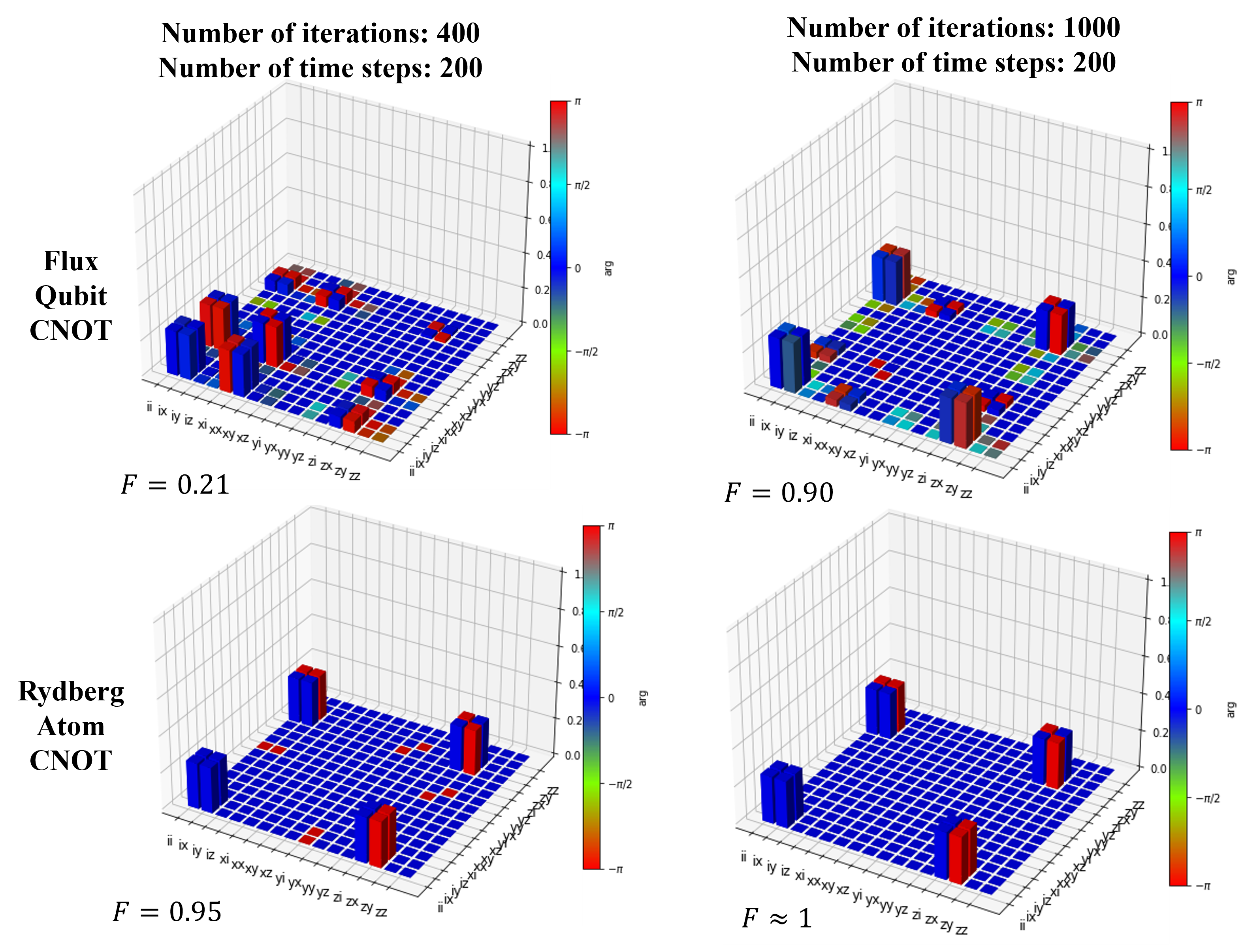}}
    \caption{Quantum process tomography evaluating the construction of the CNOT gate for both the flux (with C-shunt factor $\zeta = 10$) and Rydberg systems (each depicted on a row) against GRAPE iterations and time step. Each constructed gate contains the calculated fidelity and the axes represent $i$ as the identity $I$ and $x,y,z$ as the Pauli $\sigma_x,\sigma_y,\sigma_z$, respectively. The notation $ii$ refers to a tensor product $I \otimes I$.}
    \label{fig:cnot_gate_noise}
\end{figure}

Since the control Hamiltonians of the GRAPE algorithms were given as Pauli matrices, the algorithm is capable of correcting two-qubit and single-qubit gates. This can be done practically by modifying the amount of iterations and number of time steps the algorithm utilizes to reach a desired state. Fig. \ref{fig:cnot_gate_noise}, shows the quantum state tomography for the Rydberg atom system and the flux qubit system when running the algorithm for a different number of time and iteration steps. It is good to note that the CNOT gate is simply $\frac{1}{2}(I \otimes I + I \otimes \sigma_x + \sigma_z \otimes I - Z \otimes X)$, furthermore, an in depth study of the use of the GRAPE algorithm in quantum state tomography can be found in \cite{juan_supp}.

As can be seen in Fig. \ref{fig:cnot_gate_noise}, the flux qubit system performs poorly. For this reason, different shunt values for the shunt factor $\zeta$ will be used to observe how the system, combined with GRAPE, performs at executing the distributed phase estimation algorithm.

\subsection{Methodology \& Results}

The Hamiltonians and Lindbladians were simulated using the master equation with GRAPE-optimized gates for a system comprising two flux qubits, one storing the state and the other serving as a communication qubit, as well as four Rydberg atom qubits, where one functions as a communication qubit and the remaining three as counting qubits for the distributed phase estimation algorithm. Both the time evolution and GRAPE optimization were numerically executed on QuTiP \cite{qutip_package}. In this framework, it is assumed that there are no delays in between operations because the gates are optimized for the quantum state outputted by the last operation. A flowchart of the program and snippets of the code can be found in \cite{juan_supp}.

In order to measure the accuracy of the phase estimation algorithm, the initial state was set to an arbitrary $\varphi = 3/16$ in which the goal is to measure $U$ back where $U\ket{\psi} = e^{\varphi}\ket{\psi}$. In order to observe the significance of GRAPE on optimal control, it is assumed that the $E_2$ gate possesses perfect fidelity. The algorithm is ran with time step iteration ranges of 50-200 and 100-800 GRAPE iterations. Furthermore, each iteration was run 10 times to simulate 10 shots on a quantum computer. The probability of measuring the $\varphi = 3/16$ state was calculated by dividing the correct estimations over the total number of shots. The range was chosen to optimize for computational resources while pin-pointing an interesting results range as each experiment took more than 22 hours to execute.

\begin{figure}[t]
    \centerline{\includegraphics[width=0.4\textwidth]{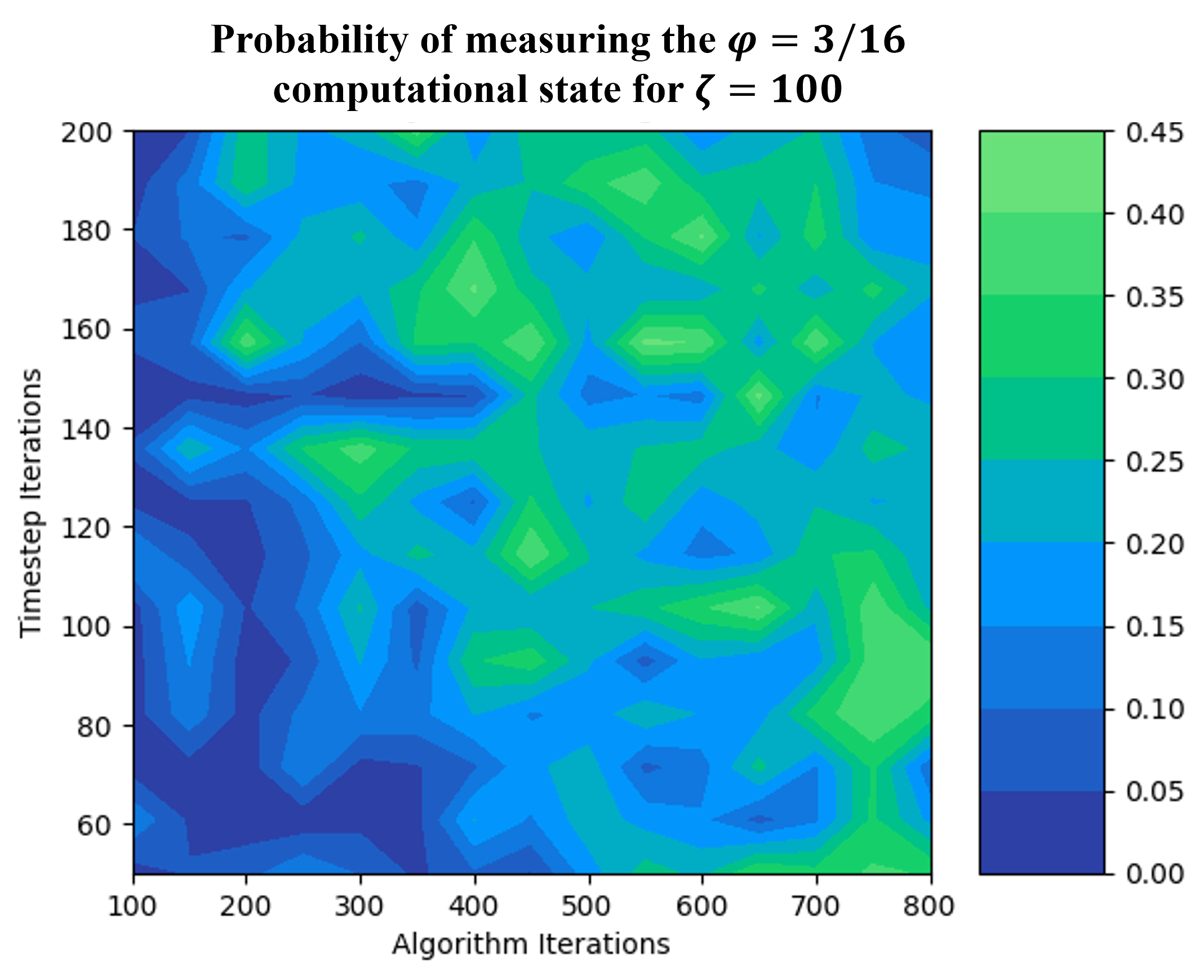}}
    \caption{Probability of measuring the $\varphi =3/16$ or $\ket{0011}$ computational state after 10 executions of the phase estimation algorithm on the noisy hybrid circuit for ranging GRAPE iterations and time steps for a shunt factor of $\zeta = $ 100.}
    \label{fig:zeta_100}
\end{figure}

\begin{figure}[b]
    \centerline{\includegraphics[width=0.4\textwidth]{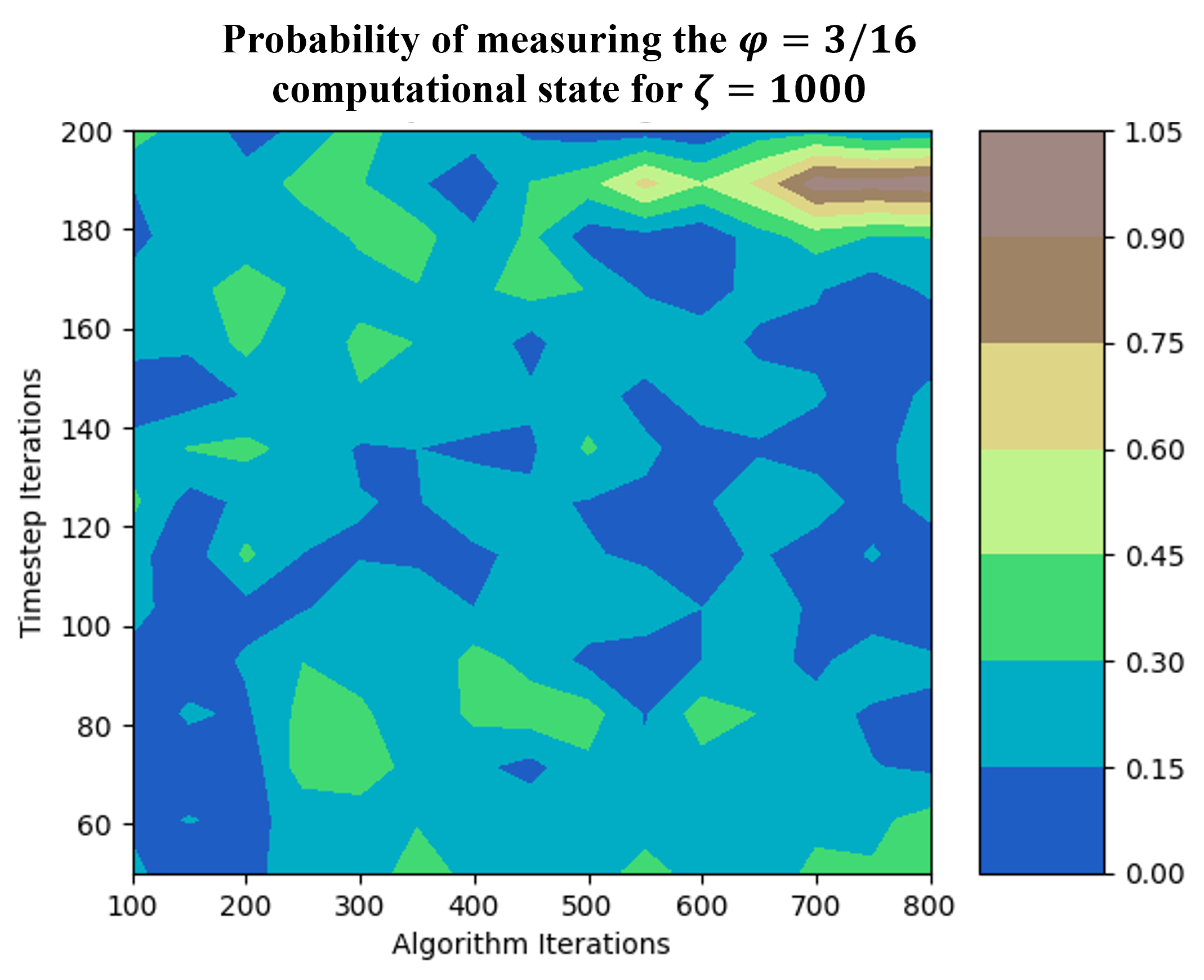}}
    \caption{ Identical concept as Fig.\ref{fig:zeta_100} but for a shunt factor of $\zeta = $ 1000.}
    \label{fig:zeta_1000}
\end{figure}

Denoted in Fig. \ref{fig:zeta_100} and \ref{fig:zeta_1000} are the results of the simulations conducted utilizing a flux qubit C-shunt factor of $\zeta = 100$ and $\zeta =1000$ respectively. It is observed that the accuracy for $\zeta=100$ and below does not improve significantly for large GRAPE iteration numbers and time step numbers \cite{juan_supp}. At this effective shunt factor, it is still not able to accurately estimate the phase for the given GRAPE iteration and time step number range. For $\zeta = 1000$, it can be observed that for approximately 190 time steps and approximately 700 GRAPE iterations, the algorithm results in an accuracy of around 95\%. The accuracy makes a steep jump from 40\% to 95\% after an increase of only 5 time steps (180 to 185 time steps). Observing the very low probability range ($<10\%$) between 700-800 GRAPE iterations and 100-160 time steps, it is noticeable that there is a trade-off between the time-step number and GRAPE iteration number occurring. 


\begin{figure}[t]
    \centerline{\includegraphics[width=0.5\textwidth]{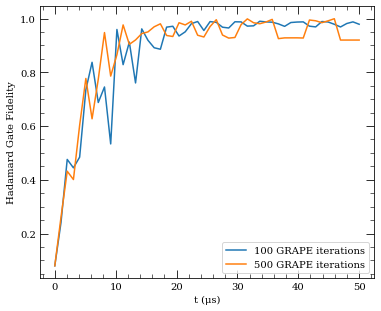}}
    \caption{Graph depicting the fidelity of a Hadamard gate for the flux qubit over time for two different GRAPE optimisation configurations. The orange line shows the evolution of the gate with 500 GRAPE iterations while the blue line shows the evolution of the gate with 100 GRAPE iterations. The 100 GRAPE iteration evolution finished with a fidelity of 0.97 while the blue line finished with a fidelity of 0.92.}
    \label{fig:grape_trade_off}
\end{figure}

To describe this trade off, the evolution of flux qubit system aiming to create a Hadamard gate can be observed using two different GRAPE configurations. One configuration is set to use 100 GRAPE iterations while the other configuration is set to use 500 GRAPE iterations. This can be seen in figure \ref{fig:grape_trade_off}. In the 100 GRAPE iteration case, the fidelity reached at the end of the evolution is 0.97 while for the 500 GRAPE iteration case, the fidelity reached is 0.92. The graph shows that, although the final fidelity for the 500 GRAPE iteration case was lower than the 100 GRAPE iteration case at the end of the run, the 500 grape iteration evolution manages to achieve fidelities higher than the 100 GRAPE iteration evolution in the total run. Observing the 500 GRAPE iteration evolution after 30 $\mu$s, one can see that the evolution holds the fidelity around 0.98 but then dips to 0.92. This dip is most likely due to the fact that the system encounters a control landscape whereby the time step required to modify the control amplitudes is not small enough to converge. Thus, at these time steps, a minimum local fidelity is reached until the landscape changes over time as seen at the 40 $\mu$s mark. This concept is further explained in \cite{Larocca_2021}. Observing figure \ref{fig:zeta_1000}, one can see that at 750 GRAPE iterations, the optimization goes from poor convergence in the control landscape at 170 time steps to proper convergence at 185 time steps. 

\section{Conclusion}
This research shows numerically that it is possible to perform the non-local phase estimation algorithm with an hybrid Rydberg-flux qubit system coupled via a co-planar waveguide. Although the fidelity of the gates in the flux system was not very high, it was shown that by engineering the qubit to possess higher C-shunt factors as well as engineering the pulse dynamics with GRAPE, one can execute the phase estimation algorithm successfully. 

The non-local GHZ state from the $E_2$ gate was theorized, but numerical simulations showed 4\% lower fidelity than prior work, likely due to simulation methods \cite{Yu_atom_flux_2017}. The non-local gates were constructed with local assumptions, neglecting asynchronicity, a key issue in distributed quantum computing when operations are not strictly sequential. The $E_2$ gate was assumed to involve instant classical communication, which is unrealistic under current classical protocols. TCP/IP networks require additional quantum operations, potentially exceeding nanosecond-scale gate creation times \cite{vanbrandwijk2016asynchronous}. Future work should explore better system couplings for higher fidelity GHZ states and address classical communication constraints.

For the flux qubit and Rydberg atom system to couple effectively, the anharmonic energy levels in Fig. \ref{fig:flux_system}(d) are necessary to avoid unintended interactions with the flux qubit system. However, C-shunt flux qubits with $\zeta = 1000$ are challenging to engineer, and they perform best with $\alpha < 0.5$, maintaining a decoherence of approximately 100 MHz \cite{juan_supp}. In this system, reducing $\alpha$ below 0.5 significantly decreases anharmonicity, causing the $\ket{R}$ and $\ket{L}$ states to vanish, forcing constant coupling to the resonator. This collapses the system into $\ket{0}$ and $\ket{1}$ states, preventing GHZ state construction. Future exploration of coupling methods that are less dependent on large anharmonicities could enable high-fidelity C-shunted flux qubits at $\alpha < 0.5$. Additionally, exploring quantum architectures with more efficient coupling techniques remains a novel area and could yield promising advancements in future research.

Although the GRAPE algorithm was primarily used in this study, other optimal control methods exist, notably Krotov \cite{Krotov} and CRAB \cite{Doria_2011}. Krotov was tested but proved too computationally expensive for multi-qubit gate optimization, though it excels at minimizing fidelity errors in large-qubit systems \cite{Krotov}. CRAB, being gradient-free, failed to converge at the required fidelity. \cite{Krotov} states that GRAPE is ideal when control parameters are discrete and have known Hermitian derivatives, whereas Krotov suits near-continuous controls without Hermitian constraints. Given these factors, GRAPE was the optimal choice for this research. Further works may consider the use of variational quantum algorithms (VQA) for error correction instead of optimal control.  

\begin{acknowledgments}

I sincerely thank my supervisor, Prof. E.J.D. Vredenbregt from the group Coherence and Quantum Technology in the Applied Physics department at the University of Eindhoven, for his guidance and invaluable feedback throughout my thesis.

\end{acknowledgments}

\bibliography{bib}
\bibliographystyle{ieeetr}

\end{document}